\documentclass[preprint,pra,aps,amsmath,epsfig]{revtex4} % double spaced
\usepackage{epsfig}
\usepackage{amssymb}
\usepackage{graphicx}

\begin{document}

%\nopagebreak{
%\samepage{

\title{Resonant Phenomena in  Antihydrogen-Hydrogen Scattering}
\author{A. Yu. Voronin$^1$  and P. Froelich$^2$}

\affiliation{$^1$ P.N. Lebedev Physical Institute, 53 Leninsky
prospect, 117924 Moscow, Russia \\
$^2$Department of Quantum Chemistry, Uppsala University,
Box 518, SE-75120 Uppsala, Sweden.}

\begin{abstract}

We present a  treatment of cold  hydrogen-antihydrogen collisions
based on the asymptotic properties of atom-antiatom interactions.
We derive general  formulas for the elastic and inelastic cross
sections and for the scattering lengths and analyze their
sensitivity to the   parameters characterizing the inelasticity of
the collision process.   Given the inelasticity, we obtain bounds
for the complex scattering length. We  investigate the influence of
strong nuclear forces and the isotope effects in $\bar{\rm H}{\rm
H}$ and $\bar{\rm H}{\rm D}$ collisions and demonstrate enhancement
of these effects due to the presence of the near-threshold narrow
${\rm H}\bar{\rm H}$ ($\bar{\rm H}{\rm D}$) states. The values of the elastic
and inelastic cross-sections with simultaneous account of
rearrangement and strong forces are presented. General
expressions for the (complex) energies of the  near-threshold
$\rm{H}\bar{\rm H}$ states are obtained.
\end{abstract}

\maketitle

{\small
\vskip 1cm
%\tableofcontents
}

\section{Introduction}

A very important property of ultra-cold atom-antiatom collisions  is
the separation of scales, characteristic for inelastic transitions
on one hand, and van der Waals  interaction on the other hand.
Indeed, the analysis of the dynamics of Protonium ($Pn$) and
Positronium ($Ps$) formation \cite{Kolos75,
Shlyap1,vc98,froe02s,voro04,froe00,jons01,zyge04,Arm04}, which is
the dominant inelastic process in $H-\bar{H}$ cold collisions, shows
that such transitions happen mainly at internuclear separation
distances less than $R_s$, which is on  the order of the Hydrogen
Bohr radius ($r_B$). As shown by previous investigations
\cite{froe00} this distance is optimal for the overlap of the
initial channel ($H-\bar{H}$) wave function and the final state
($Pn-Ps$) wave-functions.

At  separation distances larger than $R_s$ the $H-\bar{H}$
interaction is well approximated by the one-channel adiabatic
potential $V_{ad}$ \cite{Stras04, Arm04}. Therefore the
wave-function for the relative motion of  the $H-\bar{H}$ could be
obtained from the solution of the \emph{one-channel} Schr\"{o}dinger
equation with the potential $V_{ad}(r)$,  together with a proper
boundary condition imposed at the internuclear distance $r\simeq
R_s$. Importantly, such a boundary condition (which ideally should
be derived from the solution of the four-body problem at
internuclear distances $r<R_s$) turns out to be energy independent
for sufficiently small energies of colliding $H$ and $\bar{H}$.
Indeed, as follows from the kinematics of $Pn$ and $Ps$ formation
\cite{jons01}  the energy of relative motion of $Pn$ and $Ps$ is
$\varepsilon=E+M/(2N^2)-m+m/(4n^2)$, where $E$ is the $H-\bar{H}$
c.m. collision energy, $M$ is the reduced mass of $Pn$, $m$ is the
reduced mass of Hydrogen, $N$ is the principal quantum number of
$Pn$, $n$ is the principal quantum number of $Ps$. As one can see
the minimum of $\varepsilon$ is $\varepsilon_m= 0.046$ a.u. ($E=0$,
$N=24$ and $n=1$). Therefore it is reasonable to expect that for
$H\bar-{H}$ collision energies $E\ll\varepsilon_m$ the $Pn$-$Ps$
wave-function
%and the corresponding boundary condition
would weakly depend on the  collisional energy $E$. In the following
we will be interested in cold collisions at energies $E\le 10^{-5}$ a.u.
where the S-wave contribution is dominant and  the above condition is
easily  fulfilled. Thus the influence  of rearrangement and
annihilation on the wave-function in the $H\bar{H}$ channel can be
described by the \emph{complex} boundary condition for the
logarithmic derivative of the wave-function at $R_s$, which is
energy independent in the energy domain of interest.
This boundary condition can be conveniently expressed in terms of
the short-range complex
phase-shift $\delta=\delta_R+i\delta_I$, which is connected to the
above mentioned boundary condition {\it via} the relation
\begin{equation}\label{delta}
\frac{\Phi'(R_s)}{\Phi(R_s)}=p(R_s)\cot(\delta)
\end{equation}
where
$p(R_s)=\sqrt{2M V_{ad}(R_s)}$ is a classical local momentum given
at the distance $R_s$.

In our study we will be interested in the near-threshold properties
of $H-\bar{H}$ scattering. Such properties are critically dependent
on the existence of the near-threshold singularities of the
$S$-matrix, namely the weakly bound states or resonances. It is known that
the density of the near-threshold levels is determined by the large
distance behavior of adiabatic potential $V_{ad}(r)$ \cite{Fred1}.
In our case this potential at large distances  turns into
the homogeneous van der Waals potential $-C_6/r^6$.
%The asymptotic form of the interaction
%dictates also the size of the scattering length.
The  strength of this potential
 ($C_6$) determines the  characteristic distance
$R_{vdW}=\sqrt[4]{2M C_6}\simeq 10.5$ a.u.
which, as will be shown later, gives the order of magnitude  of the
scattering length for the $H\bar{H}$ system. At the same time the variation
of the  boundary condition
(related to the short range phase-shift $\delta$, see eq. \ref{delta})
will only weakly affect the density of the near-threshold levels;  it
results mainly in shifting of the whole  spectrum of such states. It
may happen that for certain value of $\delta$ there is a state
(resonance) very close to the threshold. This results in a strong
enhancement of the cross-sections.

In this paper we will study the evolution of the scattering
observables as a function of the short-range complex phase-shift
$\delta$, thought of as a free variable. Though the "true" value of
$\delta$ could be in principle obtained from the solution of the
four-body problem, it is very instructive to get  a general
dependence of the scattering amplitude on $\delta$. Indeed, the existing
model calculations of the  low energy $H-\bar{H}$ scattering are
all restricted by  approximations in which the effect of certain decay
channels are neglected.  In terms of the here developed approach any
account of additional  channels, or generally any improvement of the optical
potential,  result in a change  of the short
range phase-shift $\delta$. We  present the universal dependence
of the scattering length on the complex short-range phase-shift
$\delta$ in a closed form and show that such  dependence manifests
resonant behavior,  connected to  the existence of narrow
near-threshold $H\bar{H}$ states. We analyze  this behavior
in terms of the near-threshold  poles of the $S$-matrix.
Such resonance phenomena can
significantly enhance the influence of "small" physical effects
(or their neglect in approximate treatments) and
make inapplicable naive perturbation approach. In particular, we  study
the influence of strong forces in the
non-relativistic $H\bar{H}$ Hamiltonian, including the shift and
splitting of the quasi-bound near-threshold $H\bar{H}$ states, as
well as the isotope effect in $D-\bar{H}$ scattering.

\section{Low energy S-state $H-\bar{H}$ scattering}\label{sect1}

We start  with the derivation of the analytical expression for the
scattering length for  $H-\bar{H}$ collisions. This derivation is
based on the matching of the WKB form of the wave-function and  the
analytical zero-energy solution of the Schr\"{o}dinger equation with
the homogeneous ($-C_6/r^6$) potential in the asymptotic region.
Such an approach was successfully used for the study of cold atomic
collisions \cite{Flamb1,Dalg1,Shlyap1}.
% however, it would be
%convenient to have main formulas at our disposal here.
%It is  useful  to
%have the main steps of the derivation of the expression for the
%scattering length at our disposal here.
Our treatment will be
extended to the atom-antiatom case, where account of inelastic
transitions is important.

Further analysis is based on the fact, that the WKB approximation is
applicable for the description of the wave-function
  in between the  short-range distance and the asymptotic
domain $R_s<r\ll R_{vdW}$:
\begin{equation}\label{PsiWKB}
\Phi(r) \sim \frac{1}{\sqrt{p(r)}}\sin(\delta+\int_{R_{s}}^r{p(r')
dr'})
\end{equation}
where $p(r)=\sqrt{2MV_{ad}(r)}$ is the classical local momentum  in
the adiabatic potential $V_{ad}(r)$, while
$\delta=\delta_R+i\delta_I$ is the complex short range phase-shift,
whose  value should be fixed by  matching with the solution of
the four-body problem at $r=R_s$.

   It is important that the distance ($R_h$) above which the adiabatic potential
   $V_{ad}(r)$ is well approximated by the homogeneous $-C_6/r^6$
potential is within the
   distance of the validity of WKB approximation, i.e. $R_h<R_{vdW}$.
   The rough estimation of $R_h$ can be obtained from the condition
   that for the distances above $R_h$ the term $C_6/r^6$ should dominate in the
   multipole expansion of $V_{ad}(r)$. This gives
   $R_h=\sqrt{C_8/C_6}\sim 5$ a.u..
   Thus there exists a matching region $R_h<r\ll R_{vdW}$, where both the
WKB approximation and the zero-energy solution of
the Schr\"{o}dinger equation with homogeneous ($-C_6/r^6$) potential
are valid approximations of the $H\bar{H}$ wave-function.

  In the vicinity of $R_h$,
where  the adiabatic potential is well reproduced by a homogeneous
potential,   the classical momentum is $p(r)=\sqrt{2MC_6/r^6}$ and
one can get the explicit $r$ dependence of the wave-function introduced in
eq. (\ref{PsiWKB}):
\begin{equation}
\Phi(r) \sim \frac{r^{3/2}}{\sqrt[4]{2MC_6}}\sin(\delta +
\Omega-\frac{\sqrt{2MC_6}}{2r^{2}})
\label{psiwkb}
\end{equation}
where we have introduced the semiclassical phase $\Omega$ defined as
\begin{equation}\label{PhaseWKB}
\Omega=\int_{R_{s}}^{\infty}{p(r) dr}.
\end{equation}
The value of $\Omega$ calculated using the adiabatic potential
$V_{ad}$ of the $H-\bar{H}$ system from the ref. \cite{stra02}
% (this potential  will
%be used throughout the paper and referred to as $V_{ad}$)
and $R_s=1$ a.u. turns out to be
$\Omega=19.383$.

The wave function given in  eq. (\ref{psiwkb}) can
be matched in the vicinity of $R_h$ with the exactly known
zero-energy wave function $\Phi_0(r)$ in a homogeneous potential
$1/r^6$ \cite{MM}, which is a linear combination of the form:
\[
\Phi_0(r) \sim \sqrt{r} \left(
J_{1/4}(\frac{\sqrt{2MC_6}}{2r^{2}})-C
Y_{1/4}(\frac{\sqrt{2MC_6}}{2r^{2}}) \right).
\]
The coefficient $C$ has  to be determined from the matching
procedure at $R\simeq R_h$ which results in the following wave-function
at distances $r>R_h$:
\begin{equation}\label{Phi0}
\Phi(r>R_h) \sim \sqrt{r} \left(
J_{1/4}(\frac{\sqrt{2MC_6}}{2r^{2}})-\tan(\frac{\pi
}{8}+\Omega+\delta) Y_{1/4}(\frac{\sqrt{2MC_6}}{2r^{2}}) \right).
\end{equation}
Using the Taylor expansion of the Bessel functions for  small argument
and taking into account that the scattering length
appears in the  asymptotic form of the wave function through
\begin{equation}
\Phi(r\rightarrow \infty)\sim 1-r/a
\end{equation}
we obtain  for the scattering length
\begin{equation}\label{a6}
a =%
a_0\left( 1+\cot(\frac{\pi}{8}+\Omega+\delta)\right)%
\end{equation}
where $a_0$ is

\begin{equation} \label{a0}
a_0=R_{vdW}\frac{\Gamma (3/4)}{2\sqrt{2}\Gamma (5/4)}\simeq
4.99\mbox{ a.u.}.
\end{equation}
As one can see $R_{vdW}$ gives the characteristic size scale for the
$H-\bar{H}$ scattering length.
  The argument of cotangent in (\ref{a6}) has a simple meaning. Indeed, $\Omega$ is the semiclassical
 phase accumulated in the region $r>R_s$,
 $\delta$ is the  short range phase-shift accumulated at $r<R_s$   and the term  ($\pi/8$)   is
the quantum correction to the semiclassical phase  from the
asymptotic van der Waals tail ($-C_6/r^6$).

The accuracy of the above expression is limited by the accuracy of
the WKB approximation in the range $R_s<r<R_{vdW}$  and on
possible discrepancies between  the adiabatic potential $V_{ad}$ and
the van der Waals term $-C_6/r^6$ at the upper limit of that range.
The comparison of (\ref{a6}) with the numerical calculation of the
Schr\"{o}dinger equation with the adiabatic potential $V_{ad}(r)$  and
boundary condition (\ref{delta}) shows that formula (\ref{a6}) has
the accuracy of 10 \% in the wide range of $\delta$.

%%%%%%%%%%%%%%%%%%%Effective range %%%%%%%%%%%%%%%%%%%%%%%%%%%%%%%%%%%%
One can extend the above analysis to the calculation of the effective range. According to \cite{Gribakin2,Gao1} the
effective range is:
\[
r_e=\frac{a_0}{3}\left[\frac{\Gamma(1/4)}{\Gamma(3/4)}\right]^2\left(1-2\frac{a_0}{a}+2\frac{a_0^2}{a^2}\right).
\]
Upon substitution of $a$ from eq. (\ref{a6}) one gets
\begin{equation}
\label{effrange}
r_e=\frac{a_0}{3}\left[\frac{\Gamma(1/4)}{\Gamma(3/4)}\right]^2\frac{(\cot(\pi/8+\Omega+\delta))^2+1}{(\cot(\pi/8+\Omega+\delta)+1)^2}.
\end{equation}

  % The expression (\ref{as}) is general  and allows  study  of
%various situations
  % through the appropriate choice of $\delta_R$, $\delta_I$  and $\Omega_s$.
  % This convenient parametrization of the scattering length becomes exact for
%exact values of $\delta_R$, $\delta_I$ and $\Omega_s$.
% which could
%be obtained  via solution of the four-body problem.
For sufficiently low energy, \,i.e.  when the momentum of the incident atom
  satisfies $k|a|\ll 1$ the
  elastic ($\sigma_{el}$) and inelastic
($\sigma_{in}$) cross-sections are determined by the scattering
length:
\begin{eqnarray}
\label{CrSec} \sigma_{el}&=& 4\pi |a|^2=4\pi
a_0^2\left[(1+\mathop{\rm
Re}\cot(\frac{\pi}{8}+\Omega+\delta))^2+(\mathop{\rm
Im}\cot(\frac{\pi}{8}+\Omega+\delta))^2 \right],\\
\sigma_{in}&=& 4\pi |\mathop{\rm Im} a|/k =
4\pi\frac{a_0}{k}\mathop{\rm Im}\cot(\frac{\pi}{8}+\Omega+\delta).
\label{CrAn}
\end{eqnarray}
These values can be corrected to account for the second order terms in momentum $k$:
\begin{eqnarray}
\label{CrSec2} \sigma_{el}&=& \frac{4\pi |a|^2}{1-2k\mathop{\rm Im}a+k^2 (|a|^2-\mathop{\rm Re}(r_e a))},\\
\sigma_{in}&=& \frac{4\pi}{k}\frac{ |\mathop{\rm Im} a|}{1-2k\mathop{\rm Im}a+k^2 (|a|^2-\mathop{\rm Re}(r_e a))}.
\label{CrAn2}
\end{eqnarray}
Since   the formation of Protonium and Positronium in the
rearrangement collisions ends in complete  annihilation,  the total
inelastic cross-section can be identified with  the cross-section for
annihilation.

 We will now turn to the study of the general dependence of
the scattering lengths and cross-sections (elastic and inelastic) on
the variation of the real ($\delta_R$) and imaginary ($\delta_I$)
parts of the short range phase-shift $\delta$.

Let us treat two  important limiting cases. The first, which we
refer to as  the \emph{"weak absorption limit"}, is characterized by
small imaginary phase-shift $\delta_I\ll 1$ which  means that the
probability of transitions to the inelastic channels is  small. In
that  case we get for the scattering length $a$ from eq. (\ref{a6})
\begin{equation}\label{a6osc}
   a= a_0%
\left( 1+\cot(\frac{\pi}{8}+\Omega+\delta_R)\right)- ia_0
\delta_I/\sin^2(\frac{\pi}{8}+\Omega+\delta_R),
\end{equation}
\begin{eqnarray}
\label{CrSecW} \sigma_{el}&=& 4\pi |a|^2=2\pi a_0^2%
\frac{\sin^2(\frac{3\pi}{8}+\Omega+\delta_R)}%
{\sin^2(\frac{\pi}{8}+\Omega+\delta_R)}, \\
\sigma_{in}&=& 4\pi |\mathop{\rm Im} a|/k =
4\pi\frac{a_0}{k}\frac{\delta_I}{\sin^2(\frac{\pi}{8}+\Omega+\delta_R)}.
\label{CrAnW}
\end{eqnarray}
%
%where  $k$ is the momentum of incident atoms.

% Since our treatment assumes the short- and middle-range
%phase-shifts (contributing to
%$\alpha_6$)  to be parameters of the model,
% it is desirable to study the dependence of the observables on such
%parameters.
% An important feature of
The scattering cross sections (eqs. \ref{CrSecW}-\ref{CrAnW})
   show  oscillating character as a function of
   $\delta_R$. As will become apparent later  such oscillations
   originate from the existence of a spectrum of  long-lived  near-threshold
   states of the $H\bar{H}$ system.

   For the application to  collisional cooling we are  interested
   in the competition between the rate of elastic scattering and the rate of
   annihilation.
   The ratio $\sigma_{el}/ \sigma_{in}$ reaches  its
   maximum
\begin{equation}\label{Max}
\left( \frac{\sigma_{el}}{\sigma_{in}} \right)_{max}=\frac{a_0 k
}{2\delta_I }
\end{equation}
for $\frac{\pi}{8}+\Omega+\delta_R=\pi/4+\pi v$, $v=0,1,...$ and its
minimum
\begin{equation}\label{Min}
\left(\frac{\sigma_{el}}{\sigma_{in}}\right)_{min}=2 k a_0 \delta_I
\end{equation}
for $\frac{\pi}{8}+\Omega+\delta_R=3\pi/4+\pi v$, $v=0,1,...$ .
%Let us note, that for the
%validity of the scattering length approximation it is needed that
%$k\sqrt[4]{2MC_6} \ll 1$. It means that for the realistic values of
%$\delta_I$ one can expect that the maximum value of this ratio is
%not higher than unit in the range of the scattering length
%approximation validity.

Let us now turn to
the opposite limit of \emph{strong absorption}, characterized by
   $\delta_I\gg 1$. In this case we get
\begin{eqnarray}
\label{a6inf} a &=&a_0 (1-i),\\
\label{rinf} r_e&=&0,\\
\label{sigmael}\sigma_{el}&=&8\pi a_0^2,\\
\label{sigmainel}\sigma_{in}&=& 4\pi \frac{a_0}{k}.
\end{eqnarray}
The above  cross-sections (eqs. \ref{sigmael}-\ref{sigmainel})
are determined by the van der Waals  tail
of the adiabatic potential only. Remarkably, they
   include \emph{no information} about the short- and middle-range parts of
   the atom-antiatom interaction. All  particles that
   penetrate to these short distances are lost, so the only
   information available to the observer is due to the so called quantum
reflection \cite{cftQr1,vfzHbarWall}  from  the
   asymptotic  tail of the potential, which is determined by $C_6$
   alone.
   A remarkable feature of the strong absorption limit is that the effective range (\ref{effrange}) is exactly zero.
%%%%%%%%%%%%%%%%%%%%%%%%%%%%%%%%%%%%%%%%%%%%%%%%%%%%%%%%%%%%%%%%%%%%%%%%
\begin{figure}
   \centering
\includegraphics[width=85mm]{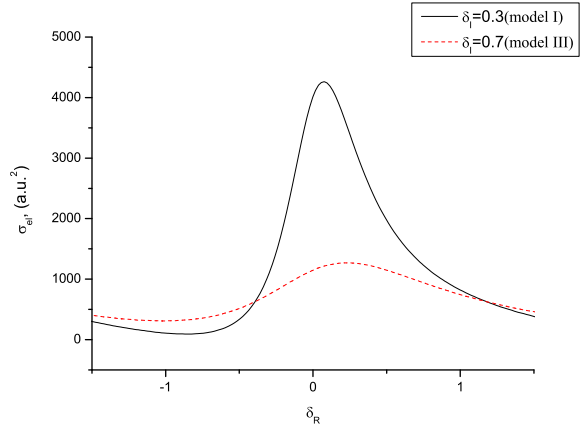}
   \caption{Elastic  cross-section for  $H\bar-{H}$ collisions as a function of phase
   $\delta_R$, calculated at $E \rightarrow 0$ according to
   $\sigma_{el} = 4\pi |a|^2$ with $a$ given by eq. (\ref{a6}). } \label{Fig1}
\end{figure}
%%%%%%%%%%%%%%%%%%%%%%%%%%%%%%%%%%%%%%%%%%%%%%%%%%%%%%%%%%%%%%%%%%%%%%%%%%%
%%%%%%%%%%%%%%%%%%%%%%%%%%%%%%%%%%%%%%%%%%%%%%%%%%%%%%%%%%%%%%%%%%%%%%%%
\begin{figure}
   \centering
\includegraphics[width=85mm]{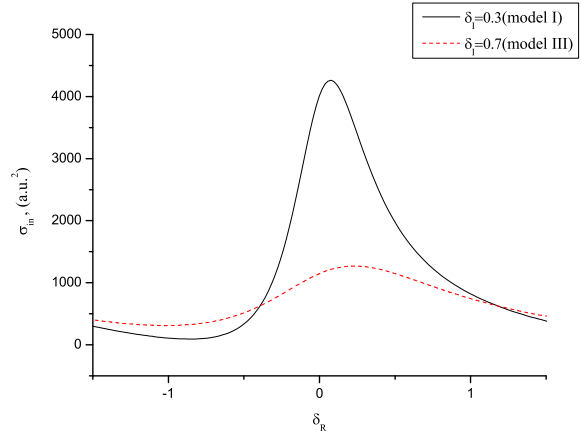}
   \caption{Annihilation  cross-section for  $H-\bar{H}$ collisions as a function of
   phase $\delta_R$, calculated at the energy E=$10^{-6}$ a.u.
   according to
   $\sigma_{in} = 4\pi Im(a)/k$ with $a$ given by eq. (\ref{a6}).}
   \label{Fig2}
\end{figure}
%%%%%%%%%%%%%%%%%%%%%%%%%%%%%%%%%%%%%%%%%%%%%%%%%%%%%%%%%%%%%%%%%%%%%%%%%%%
%%%%%%%%%%%%%%%%%%%%%%%%%%%%%%%%%%%%%%%%%%%%%%%%%%%%%%%%%%%%%%%%%%%%%%%%
\begin{figure}
   \centering
\includegraphics[width=85mm]{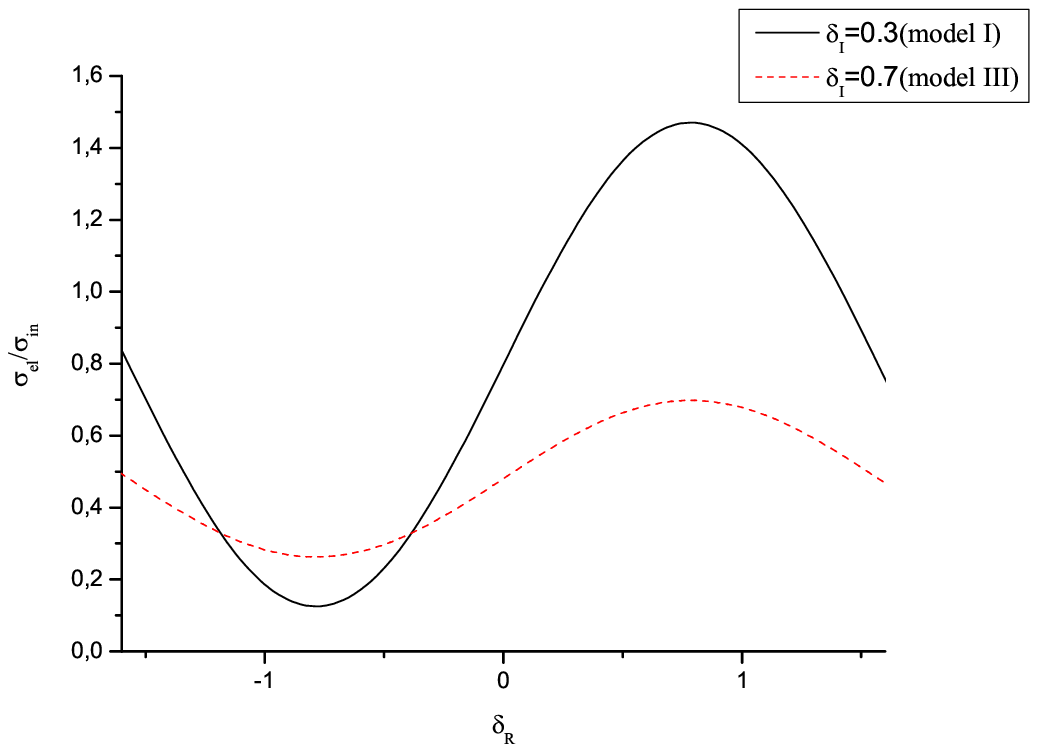}
   \caption{Ratio of the elastic to the annihilation cross-section as a
   function of phase $\delta_R$, calculated at the energy E=$10^{-6}$ a.u.
    with $a$ given by eq. (\ref{a6}).
   }\label{Fig3}
\end{figure}
%%%%%%%%%%%%%%%%%%%%%%%%%%%%%%%%%%%%%%%%%%%%%%%%%%%%%%%%%%%%%%%%%%%%%%%%%%%

In  Fig. \ref{Fig1} we plot the elastic cross-section as a function
of real part of short-range phase-shift $\delta_R$ for two values of
the "inelasticity parameter" $\delta_I=0.3$ and $\delta_I=0.7$.
Pronounced oscillations can be seen in the "weak absorption" case,
which become much more smooth with increasing inelasticity parameter
(in the limit $\delta_I \gg 1$ the cross-sections are constant). The
same tendency is apparent in the inelastic cross-section (Fig.
\ref{Fig2})  and in the ratio of the elastic to the annihilation
cross-section (Fig. \ref{Fig3}).

   \section{Physical implications}\label{sect3}

   In this section we discuss the physical implications  of
   the  sensitive  dependence of the scattering length on the short-range phase
   shift. We   study   the evolution  of
   the scattering length as  function of the phase-shift
   $\delta_R$. This can be considered as an  universal tool for
   studying the contribution
   of various  physical effects to  scattering
   observables.

   \subsubsection{Limits on the accuracy of the model $H_{1S}-\bar{H}_{1S}$
   calculations}

Since the complete    \textit{ab initio} treatment  of the
hydrogen-antihydrogen scattering
   is a  difficult problem
   it  is  important
   to be able to estimate the accuracy of the approximate    calculations.
   Several model calculations have been performed to get elastic
   and annihilation cross-sections in
   the case of $H_{1S}-\bar{H}_{1S}$ scattering
\cite{jons01,froe02s,froe04,jons04,zyge04,armo02,armo03,sinh00,sinh03,vc98,voro98,voro01,voro04,labzw}.
    In these models mostly the same
   adiabatic  potential $V_{ad}$ of $H_{1S}-\bar{H}_{1S}$ was used,
applicable in the distance
   range $r>R_s\simeq 1 r_B$ \cite{stra02}.
  At  the same
   time very different approximations were made regarding the
rearrangement and strong
   forces. They include  calculations of the elastic cross-section
   neglecting both rearrangement and strong force \cite{jons01}, the account of
rearrangement in the
   "distorted-wave" approximation \cite{jons00, froe00, jons04}, calculation of the annihilation
cross-section due to
   strong force only \cite{jons04,armo03},  non-perturbative calculations
based on the
approximative optical
   potential \cite{zyge04,voro04} and other.

    In terms of the formalism developed in the previous section
 the above
   mentioned models differ by the short-range phase $\delta=\delta_R +
   \delta_I$, i.e. the phase accumulated up to $R \simeq R_s$.
This phase incorporates the effects of inelasticity due to strong
forces and rearrangement. In the  optical model calculations by Voronin and Carbonel \cite{voro04}
   (further referred to as model I) it was found that $a=5.2-i 1.8$ a.u.
   (without account of strong force).
   From this value of the scattering length the corresponding value of the short-range
   phase-shift can be uniquely deduced (in the sense that the solution of
   the Schr\"{o}dinger equation with given adiabatic potential $V_{ad}$ and with boundary condition (\ref{delta})  results in
   unique correspondence between $\delta$ and $a$, as  can be seen from
eq. (\ref{a6})). The value of the short range phase-shift
   for model I turns out to be
   $\delta=0.696+i 0.305$. The value presented in  \cite{armo02} by  Armour,
Liu and  Vigier (further referred to as model II) is $a=8.2-i2.8$ a.u.
and the short-range phase-shift deduced from this value of the
scattering length is $\delta=0.136+i0.41$. The optical model
calculation \cite{zyge04} by  Zygelman, Saenz, Froelich and Jonsell
(further referred to as model III) gives $a=5.6-i3.7$ a.u. (without
strong force). The corresponding short-range phase-shift turns out
to be $\delta=0.638+i0.715$.   Finally, the calculation
of  elastic scattering
neglecting   both rearrangement
and strong force effects (further referred to as model
IV, \cite{jons00}) results in the scattering length
$a=7.69$ a.u..  The corresponding \emph{real} phase-shift, deduced
from this value of the scattering length is $\delta=0.287$.
All these values are tabulated in Table \ref{TableI}.
Inclusion of the inelasticity is thus equivalent to introduction of the
short-phase corrections to model IV. Whereas it was possible to
separately calculate
the complex phase due to direct annihilation \cite{jons04,jons04a},
calculation of such
phase correction due to the rearrangement is still incomplete since
it is based only on
the imaginary component of the optical potential \cite{zyge04}.
Hence the inaccuracy of
the real phase $\delta_R$ remains an important source of the inaccuracy of
the scattering length in $H-\bar{H}$ scattering.

 From the eqs. (\ref{CrSecW},\ref{CrAnW}) it
   is clear that the smaller  is the inelasticity parameter $\delta_I$,
 the more
   sensitive are the model results to the  uncertainty in the
   {\it real }
   part of the $H-\bar{H}$ interaction at distances $r<R_s$.
% (which is responsible for uncertainty in $\delta_R$).
We notice that for a
  fixed value of inelasticity parameter $\delta_I$ the relation
  between imaginary and real part of the scattering length can not be
  arbitrary but  is determined by eq. (\ref{a6}). To illustrate this
  statement we plot the possible values of the  scattering length
  $a$ in the complex plane of $a$
  as a function of $\delta_R$ (which changes from $-\pi/2$ to $\pi/2$)
for  fixed value of
  $\delta_I$. The possible values of $a$ for each fixed value of
  $\delta_I$ form closed curves.
%%%%%%%%%%%%%%%%%%%%%%%%%%%%%%%%%%%%%%%%%%%%%%%%%%%%%%%%%%%%%%%%%%%%%%%%
\begin{figure}
   \centering
\includegraphics[width=85mm]{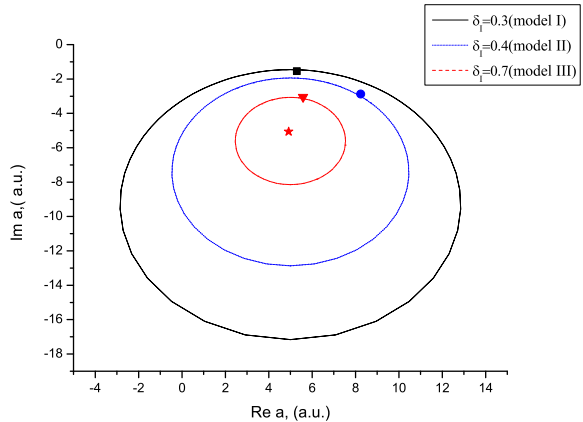}
   \caption{Complex scattering length $a$ as  function of the short-range phase-shift
$\delta_R$,
    for the fixed  value of $\delta_I$. The square indicates the  complex value
of the scattering length obtained in model I, circle - in model II, triangle - in model
    III.
    The value of $a\simeq 5-i5$ a.u. marked by a star corresponds to the limiting case of strong absorption $\delta_I\gg 1$.
   }\label{FigArg}
\end{figure}
%%%%%%%%%%%%%%%%%%%%%%%%%%%%%%%%%%%%%%%%%%%%%%%%%%%%%%%%%%%%%%%%%%%%%%%%%%%
The three curves of the scattering length  correspond to three
different inelasticity parameters $\delta_I$ deduced  from  models
I-III. One can see  that the smaller is $\delta_I$ the larger is the
possible variation of $a$ with $\delta_R$. In the limit of zero
absorption ($\delta_I\rightarrow 0$) the variation becomes infinite
and the closed curve degenerates into the whole real axis  of the
complex plane (model IV), while in the opposite limit of strong
absorbtion the curve shrinks into a point $a=a_0(1-i)\simeq 5-i5$
a.u.. Interestingly, the mean value of the scattering length
$\bar{a}$ (averaged over $\delta_R$) is the same for all curves,
i.e. it turns out to be \emph{independent} on the inelasticity parameter
$\delta_I$ and is equal to $\bar{a}=a_0 (1-i)$.

The corresponding variations of the cross-sections are shown on Fig.
\ref{Fig1} and Fig. \ref{Fig2} for models I and III (the
corresponding inelasticity parameters are $\delta_I=0.3$ and
$\delta_I=0.7$, respectively). One can see that the ratio of maximum
to minimum possible values
 of elastic cross-section turns out to be $\sigma_{el}^{max}/\sigma_{el}^{min}\simeq
 46$ for $\delta_I=0.3$. The same value for $\delta_I=0.7$ is much
 less and turns out to be $\sigma_{el}^{max}/\sigma_{el}^{min}\simeq
 4$. In view of this result it is clear that the knowledge of the
absorptive phase-shift $\delta_I$ is very essential and it is
sufficient  for establishing the limits of accuracy in  calculations
of  both  inelastic and  elastic cross-sections.

We can perform further analysis and estimate how the uncertainty
in  the  $H-\bar{H}$ interaction at distances $r<R_s$ would be reflected
in the uncertainty of the scattering observables.
Such an estimation can not be model independent. We will use the
following simple assumptions. One can expect
\cite{jons04,armo02,labzw} that the leading contribution to the
mentioned interaction at distances $R<R_s$ is given by the
  local interaction of the form (expressed in a.u.):
  \begin{equation}\label{Vsloc}
   V_{loc}(r)=-1/r+0.75.
   \end{equation}
  This potential continuously matches with  $V_{ad}(r)$ at $r\approx R_s$
  a.u. Apart from such local potential  there is an additional interaction in
the $H-\bar{H}$ channel,
  which results from  coupling to the decay channels.
We will model the effects of coupling to the  $Pn$ and $Ps$ channels,
responsible for rearrangement,  by the  local complex potential of the
form:
\begin{equation}\label{Vopt}
V_{opt}=\frac{(v-iw)}{r_B} exp(-2 r).
\end{equation}
The above simple form of "optical" potential
is only used to mimic the effect of coupling to the  decay channels and
can not be treated as true form of effective short-range
$H-\bar{H}$ interaction - but it captures its localization.
The \textit{ab initio} calculations of
complex nonlocal optical potential (so far restricted by
approximations) can be found in \cite{voro98,voro04, zyge04}. These
optical potentials are constructed by projecting  the Hamiltonian
on the subspace of open $Pn$ and $Ps$ channels and therefore  their
characteristic range is equal to the "size" of the $Pn$ states with
energetically highest possible principal quantum numbers ($N\leq
24$). Because of that such \textit{ab initio} optical potentials are
localized below $R_s$ and this property is reflected in the $r$
dependence of the simple model potential $V_{opt}(r)$.

 To calculate the short-range phase-shift we will use the semiclassical approximation for the phase:
 \begin{equation}\label{modelDelta}
 \delta\approx \int_0^{R_s} \sqrt{-2 M [V_{loc}(r)+V_{opt}(r)]}dr -\pi/4.
 \end{equation}

 The  term $\pi/4$ is the  quantum correction to account for the
 correct behavior of the coulomb wave-function  at small distances $r<1/M$, where
 semiclassical approximation is no longer valid.
Expecting $|V_{opt}(r)|\ll |V_{loc}(r)|$ we get  from
(\ref{modelDelta}) for  $\delta$:
\begin{eqnarray}
\label {deltaresult}
\delta&=&\delta^0+\delta^{opt},\\
\delta^0&=&\int_0^{R_s}\sqrt{-2 M
V_{loc}(r)}dr,\\
\delta^{opt}&=&-\sqrt{M}\int_0^{R_s}\frac{V_{opt}(r)}{\sqrt{-2V_{loc}(r)}}dr.
\end{eqnarray}
%%%%%%%%%%%%%%%%%%%%%%%%%%%%%%%%%%%%%%%%%%%%%%%%

In the above  expression $\delta^0$ is a phase-shift produced by the
local potential $V_{loc}$ alone, while $\delta^{opt}$ is the
variation of that  phase-shift due to the presence of
the optical potential. Using the adopted expressions for  $V_{loc}$ and
$V_{opt}$ we come to the following numbers: $\delta^0=72.45$,
$\delta^{opt}=-(v-iw)6.38$.

The corresponding variation of the scattering length can be obtained
from eq. (\ref{a6osc}):
\begin{equation} \label{acc}
   \delta a=-a_0\frac{ \delta^{opt}}{
   \sin^2(\pi/8+\Omega+\delta^0+\delta^{opt})}.
\end{equation}
%%%%%%%%%%%%%%%%%%%%%%%%%%%%%%%%%%%%%%%%%%%%%%%%%%%%%%%%%%%%%%%%%%%%%%%%
From the above numbers we can conclude, that the value of
inelasticity parameter $\delta_I$ consistent with models I-III
corresponds to the value of the imaginary part of the optical potential $w$
which lies in-between $0.05$ and $0.1$ a.u.

\begin{figure}
   \centering
\includegraphics[width=85mm]{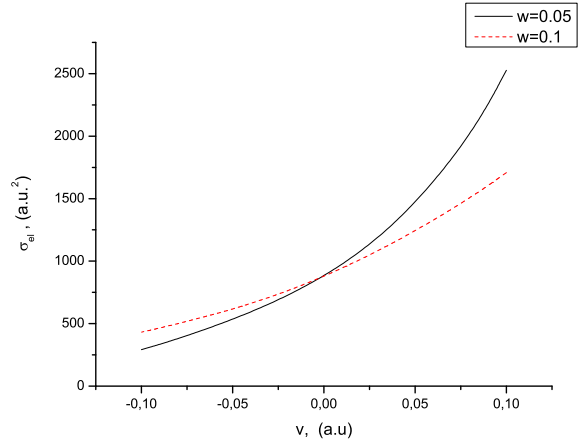}
   \caption{Elastic cross-section as a
   function of the real part  $v$ of the optical potential (\ref{Vopt}),
calculated at the energy
E=$10^{-6}$ a.u.}\label{FigOptEl}
\end{figure}

\begin{figure}
   \centering
\includegraphics[width=85mm]{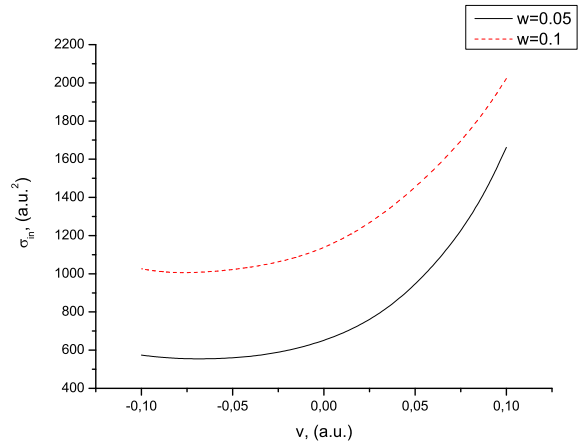}
   \caption{Inelastic cross-section as
   function of the real part  $v$ of the optical potential (\ref{Vopt}),
calculated at the energy
E=$10^{-6}$ a.u.}\label{FigOptIn}
\end{figure}

On Fig. \ref{FigOptEl}
and Fig. \ref{FigOptIn} we plot
the elastic and inelastic cross-sections
as function of the \textit{real} part of the model optical potential $V_{opt}$.
Even a small  and confined real  part  produces large variations
of the cross sections.
For the optical potential given by eq. (\ref{Vopt}),  the mean variation
of the ratio $\frac{Re V_{opt}}{V_{loc}}$
within the interval $[0,R_s]$ is
on the order of 1\%, yet it produces more than
100\% change of   elastic and inelastic scattering-cross sections. This
result makes clear the relative importance of the real part of the effective
$H-\bar{H}$ short-range interaction and puts obvious restrictions on
the accuracy of calculations. It means in particular that the
\emph{non-perturbative} account of additional decay channels could
result in significant changes of the elastic and inelastic
cross-sections. We will show in the following that such an
"instability" of  model calculations (for relatively small values
of inelasticity parameter $\delta_I$)  is due to the existence of
the spectrum of narrow near-threshold states in the $H-\bar{H}$
channel.
%%%%%%%%%%%%%%%%%%%%%%%%%%%%%%%%%%%%%%%%%%%%%%%%%%%%%%%%%%%%%%%%%%%%%%%%%%%
%%%%%%%%%%%%%%%%%%%%%%%%%%%%%%%%%%%%%%%%%%%%%%%%%%%%%%%%%%%%%%%%%%%%%%%%

   \subsubsection{Sensitivity to the nuclear interaction}

   An  important consequence of the weak absorption in
$H_{1S}-\bar{H}_{1S}$ interaction
   is the possibility of  observing  the nuclear effects on the molecular scale. The
   effect of the nuclear forces was reported in ref. \cite{jons04, armo03, voro01,voro04}
for the case of hydrogen-antihydrogen scattering  and in ref.
\cite{jons04a} for  antihydrogen-helium scattering. In calculations
\cite{jons04,armo02}  no rearrangement effects were taken into
account.

The strong force effects can be studied  by inclusion of the complex
short-range \textit{nuclear} potential in the nonrelativistic
Hamiltonian describing the $H\bar{H}$ system.
         This  potential accounts for the nuclear interaction of proton and
antiproton on the scale of few fm ($1.88 \cdot 10^{-5}$ a.u.) and is
chosen to reproduce the
  nuclear $p\bar{p}$ (spin dependent) scattering length.
Incorporation  of the   short-range complex  potential in the
   model calculations results in modification of $\delta_R$ and
   $\delta_I$. Strictly speaking, the effect of  strong forces is
not restricted to  "direct" annihilation and scattering on the nuclear
   potential in the $H-\bar{H}$ channel. It appears also through
modification of the final states and energies of Protonium.
Neglecting for the moment these "second order" effects, one can use
the well-known
  expression \cite{Watson,true61} for the nuclear phase-shifts in
the presence of
Coulomb interaction to extract directly the phase-shift of interest:
\[k\left[ C(\eta)\cot{\tilde{\delta}_{sc}}+2\eta h(\eta)\right]=-1/\tilde{a}_{sc}\]
where  $\tilde{a}_{sc}$ is the Coulomb corrected nuclear scattering length,
$\eta=-M/k$ ($k$ is the c.m. proton-antiproton momentum) and functions
$C(\eta)$, $h(\eta)$ are given by
\[C(\eta)=\frac{2\pi\eta}{\exp(2\pi\eta)-1}, \]
\[h(\eta)=1/2[\Psi(-i\eta)+\Psi(i\eta)-\ln(\eta^2)]. \]
In the above,  $\Psi(x)=\Gamma'(x)/\Gamma(x)$, where $\Gamma(x)$ is a standard
gamma-function. Neglecting the leptonic energy in comparison with
the value of Coulomb $p\bar{p}$ potential  at the distance around 1 fm,
we can put $k\rightarrow 0$ and get:
\[\tilde{\delta}_{sc}\approx-2\pi M \tilde{a}_{sc}.\]
The values of the strong-force scatering lenghth $\tilde{a}_{sc}$ in the
Kohno-Weise model \cite{KW,Carb} are:
\begin{eqnarray}
\tilde{a}_{sc}(S=0)&=&(1.07-i1.45) \cdot 10^{-5} \mbox{ a.u.}\label{ascS0},\\
\tilde{a}_{sc}(S=1)&=&(1.68-i1.06) \cdot 10^{-5} \mbox{ a.u.}\label{ascS1}
\end{eqnarray}
while the corresponding nuclear phase-shifts are
\begin{eqnarray}
\tilde{\delta}_{sc}(S=0)&=&-0.06+i0.08 \label{dscS0},\\
\tilde{\delta}_{sc}(S=1)&=&-0.1+i0.06\label{dscS1}
\end{eqnarray}
where $S=0,1$ are the values of  total nuclear spin.

The models I-IV include no information about strong forces. However
we will show that,  once the value of scattering length without any
account of strong forces is known, one can also obtain the corrected value of
this scattering length that includes the presence of strong forces. In
fact, adding the nuclear phase-shifts (\ref{dscS0}) and
(\ref{dscS1}) to the already discussed   phase-shifts deduced from the
rearrangement calculation of models I-IV
\[\delta\rightarrow\delta+\tilde{\delta}_{sc}, \]
 one gets the values of the
scattering lengths and  cross-sections, corresponding to
simultaneous account of rearrangement and strong force effects.
The above mentioned addition of  phase-shifts can be done  due
to the fact that contributions to the phase from each of the
effects (strong force and rearrangement) come from very different
distances. This is true until the effect of strong force on the
final $Pn$ states is not taken into account.

The results of our
calculations for the elastic and inelastic scattering cross-sections
with and without strong force, based on the  parameters deduced from
models I-IV are collected in  Table \ref{TableI}.
We observe that the {\it simultaneous} account of rearrangement
and strong-force effects  significantly differs from the results where only one of
these  effects is taken into account.
%%%%%%%%%%%%%%%%%%%%%%%%%%%%%%%%%%%%%%%%%%%%%%%%%%%%%%%%%%%%%%%%%%%
\begin{table}
\centering
\begin{tabular}{|c|c|c|c|c|}
  \hline
          &Model I & Model II & Model III & Model IV\\
  \hline
  $a$, a.u.     & 5.2-i1.8    &8.2-i2.8    & 5.6-i3.7    & 7.7-i0\\
  \hline
  $a^{sc}(S=0)$,a.u. &  5.6-i2.2    &8.3-i3.4   & 5.8-i4.0    &8.4-i0.4\\
  \hline
  $a^{sc}(S=1)$, a.u. &  5.8-i2.1   &  8.6-i3.4 &  6.0-i3.9   &8.1-i0.6\\
  \hline
  $\delta$ & 0.696+i0.305 & 0.136+i0.41& 0.638+i0.715&0.287+i0  \\

  \hline
  $\sigma_{el}$, a.u.$^2$ &380 & 943 & 566 & 745\\
  \hline
  $\sigma_{el}^{sc}(S=0)$, a.u.$^2$&448 & 1018 & 623& 889 \\
  \hline
  $\sigma_{el}^{sc}(S=1)$, a.u.$^2$ &472& 1077 & 640& 829  \\
  \hline
  $\sqrt{E}\sigma_{in}$, a.u. & 0.53&0.8 & 1.08 & 0\\
  \hline
  $\sqrt{E}\sigma_{in}^{sc}(S=0)$, a.u.&0.65&1.0 & 1.17&0.13 \\
  \hline
  $\sqrt{E}\sigma_{in}^{sc}(S=1)$, a.u.&0.62&1.0 & 1.15 &0.18\\

  \hline

\end{tabular}
\caption{ Values of the short-range scattering lengths,
  phase-shifts and cross-sections (in a.u.$^2$) calculated at the
energy $E=10^{-6}$ a.u.   with   the parameters deduced from different
models I-IV for $H-\bar{H}$ interaction. The  superscript $sc$
corresponds to the account of the  strong force in the presence of Coulomb
field. The  results were obtained
by  numerical solution  of the one channel Schr\"odinger equation
subject to the boundary condition given by eq. \ref{delta} with the
phase $\delta$ from row 5 incremented by the strong force contribution
given by eqs. (\ref{dscS0},\,\ref{dscS1}).}
  \label{TableI}
 \end{table}
%%%%%%%%%%%%%%%%%%%%%%%%%%%%%%%%%%%%%%%%%%%%%%%%%%%%%%%%%%%%%%%%%%%%%%%

The modification of  cross-sections due to the strong force is
 significant when compared  to the  ratio of nuclear
 and  atomic scales, which is on the order of $10^{-5}$.
 One can also see that this  modification is model dependent.
This fact is clear from our
 previous considerations of the  cross-sections as function of the
 the short-range phase-shift. In particular,
 the change  of the elastic cross-section  due to the strong force
 is around 24\% in model I, while it is 13\% in  model III. These models
 differ by the inelasticity parameter $\delta_I$ which  is
 $2.3$ times larger  in model III as compared to  model I.
Consequently  model III
 is less sensitive to any variation of phase,
in particular the phase induced by strong interaction.
The physical reason for the lack of sensitivity to the
details of the short-range interaction with the increase of
inelasticity parameter is clear. Indeed, the amplitude of the wave
reflected back into the $H-\bar{H}$ channel that is generated at such
small distances and "carries back" information about the nuclear
forces is, in the case of strong absorption, exponentially small.
Hence in the case of strong absorption  the scattering length  only
weakly depends on the details of the short range interaction.

Let us mention here that the effect of nuclear forces on $H-\bar{H}$
interaction would be strongly enhanced in the vicinity of certain
value of the real part of the short-range phase. According to
(\ref{a6osc}) the largest variation of the scattering length with
small variation of the phase $\delta_R$ is achieved when the
following condition is fulfilled

\begin{equation}\label{resonance}
\pi/8+\Omega+\delta_R=\pi v
\end{equation}
where $v$ is an integer  number.
As shown later in sec. \ref{sec:qms}   this
condition corresponds to the
appearance of the loosely bound $H\bar{H}$ state. In particular,
for the
inelasticity parameter of model I ($\delta_I=0.3$) the modification
of the elastic cross-section due to the strong force in the
near-resonance situation (eq. \ref{resonance}) would be from
$\sigma_{el}=3800$ a.u.$^2$ without strong force to
$\sigma_{el}^{sc}(S=1)=2015$ a.u.$^2$ and
$\sigma_{el}^{sc}(S=0)=2249$ with strong force, i.e. around 50\%.

%This result could be interesting in view of possibility of "fine
%tuning" of cold atoms interactions with magnetic or electric fields
%\cite{AtomFesh1,AtomFesh2,Tuning,MYElectr} in order to achieve the
%resonant condition.

   \subsubsection{Isotope effect}

The   oscillatory  behavior
   of the cross sections gives rise
   to   the  strong isotope effect. Simple changing of the
   reduced mass alters   the phases $\delta$ and $\Omega$ in eq.  (\ref{a6osc}).
The mass dependence of the phases can be easily established in the WKB
approximation (\ref{modelDelta}), from which it follows that the
semiclassical phase is scaled as the square root of the reduced mass
$M$. The change of the reduced mass from $ M_{p}/2$ to $2/3 M_{p}$
(where $M_p$ is the proton mass) corresponds to the replacement of
hydrogen by deuterium and results in multiplication of the
semiclassical phase by the factor $\sqrt{4/3}$
\[\Omega_d= \sqrt{4/3}\Omega, \]
\[\delta_d=\sqrt{4/3}(\delta-\pi/4)+\pi/4.\]
Here the subscript $d$ refers to deuterium. In the above formula we took
into account the Coulomb quantum correction $-\pi/4$.

  The numerical calculation of the
scattering length while \emph{neglecting}  rearrangement and strong
force (i.e. using potential $V_{loc}(r)$  as the $D-\bar{H}$
interaction for $r<R_s$) gives
\[a^{D\bar{H}}=-41.07 \mbox{ a.u.} \]
to be compared with the corresponding scattering length (without
rearrangement and strong force) in the $H-\bar{H}$ case
\[a^{H\bar{H}}=7.69 \mbox{ a.u.} \]

The account of rearrangement effects dramatically changes the
situation.

   The model calculations by Voronin and Carbonell
\cite{voro04} show that the scattering length changes  from
$a=5.2.1-i1.8$ a.u. for $\bar{H}-H$ to $15.0-i11.6$ a.u.  for the
$\bar{H}-D$ case. In view of the  demonstrated  high sensitivity of
the results to the short range interaction,   very accurate
calculations are required to obtain the reliable values of the
cross-sections. However  the prediction of a strong isotope effect
has a general character and is a consequence of  \,"weak absorption" in
the $H-\bar{H}$ interaction. Note that in  the opposite case of \,"strong
absorption" the expression (\ref{a6inf}) predicts only weak
monotonous dependence of the scattering length on the reduced mass
via $a_0\sim\sqrt[4]{2MC_6}$.

   \subsubsection{ The near-threshold  quasimolecular states of $H-\bar{H}$}
\label{sec:qms}

   The  oscillating behavior of the scattering
   cross-section as a function of the phase $\delta$
   has a clear physical meaning. The long range character
   of the attractive atom-antiatom interaction combined with the weak
   absorption supports the existence of long-lived  highly excited
   $H-\bar{H}$ states \cite{vc98,voro98, zyge04, froe04}.
They manifest themselves as near-threshold
   singularities of the $S$-matrix and thus strongly affect the low
energy scattering.
By changing the phase through varying the strength of interaction,
changing the  reduced mass of the system,  including additional decay
channels, etc.,  the position of such states  can be tuned to be at
the threshold,   inducing  a rapid increase of the cross-sections as
a function of the tuning parameter.
% We will estimate the energy and width of the
   %state nearest to the threshold.
The   $S$-matrix of the finite range potential can be expressed through the position of all its
poles via the following expression \cite{Newton}:
 \begin{equation}\label{Sexact}
 S=\prod_k\frac{z_k+k}{z_k-k}exp(-2ik \bar{a })
 \end{equation}
where $z_k$ is the pole of $S$-matrix and $\bar{a}$ is
independent on $k$.

 If there is a $S$-matrix pole $z$, situated much closer to the
threshold than other poles,  we can approximate the infinite product
by one term only:
\begin{equation}
S(k)\approx\frac{z+k}{z-k}exp(-2ik \bar{a }). \label{smatrix}
\end{equation}
 In the limit
of small $k$ the expression (\ref{smatrix}) becomes $S(k)=1+2k/z-2ik
\bar{a}$, from where one gets  the connection between the scattering
length and $z$
\begin{equation}
a=i/z+\bar{a}.
\end{equation}
When there are
two (or more) poles at similar distance from the threshold (bound
and virtual states) one should take into account the contribution of
the number of equidistant   poles of the $S$-matrix so that  the expression for
the scattering lengths becomes
$a=i\sum_k 1/z_k+ \bar{a}$.

It was shown in \cite{Flamb1, Gao2} that the above expression is applicable in case of scattering on
potentials with $1/r^6$ tail. In such a case the  constant $\bar{a}$  in the
one-pole expansion  is $\bar{a}=a_0$, with $a_0$
given by eq. (\ref{a0}). Therefore  the position of the pole  in
the one-pole expansion of the $S$-matrix is  given by
\[ z=i/(a-a_0). \]
  This relation  is valid in the case when
the distance between poles is much bigger then $z$.
Substituting the expression for the scattering length given by eq.
(\ref{a6}) we   get
% \begin{equation}\label{z}
%z=-\frac{i}{\left( 2MC_{6}\right) ^{1/4}}\frac{2\Gamma (5/4)}{\Gamma (3/4)}%
%\frac{\sin(\alpha_{6}+i\delta_I)}{\sin(\frac{\pi}{4}+\alpha_{6}+i\delta_I)}%
%\end{equation}
%
\begin{equation}\label{z}
z=\frac{i}{a_0}\tan(\delta+\Omega+\pi/8).
\end{equation}
Taking into account that the near-threshold states are possible only
in case of weak absorption ($\delta_I\ll1$) we obtain
\begin{equation}\label{zclose}
z=\frac{i}{a_0}\tan(\delta_R+\Omega+\pi/8)-\frac{\delta_I}{a_0
\cos^2(\delta_R+\Omega+\pi/8)}.
\end{equation}
The above equation is valid only for the
expansion with respect to the single nearest-threshold pole,
which gives the dominant contribution
to the scattering length.

From the above equation   it is
clear  that due to the inelasticity of  scattering characterized by
$\delta_I$,  the near-threshold $S$-matrix poles are shifted to the
left from the imaginary axis of the complex $k$-plane.
These shifted poles in the 2-nd quadrant of the complex $k$-plane can be
interpreted as  localized (bound) states with an inelastic width,
while the poles in the 3-rd quadrant are the virtual states or resonances
with inelastic width. (The distinction between resonances and
virtual states can be obtained by a continuous decrease of
inelasticity. In the limit of zero inelasticity the virtual states
come to the negative imaginary axis of the $k$-plane, while the resonances
are manifested as pairs of poles in the 3-rd  and 4-th  quadrants.)

The condition for the appearance of
the new state now means that  the pole comes from the third
to the second quadrant
of the complex $k$-plane. At the moment when the
pole of the $S$-matrix crosses the real (negative) axis of the complex
k-plane one has $\mathop{\rm Im} z=0$. According to  eq.
(\ref{zclose}) this is equivalent to condition (\ref{resonance}):
\[\delta_R+\Omega+\pi/8=\pi v\]
where $v$ is an integer number. Numerical
integration of the adiabatic potential $V_{ad}$ gives the WKB phase
$\Omega=19.383$
so that the numerical
value of $\delta_R$ (by modulus $\pi$) corresponding to the
resonance is $\delta_R=-0.926$.

 We have shown that the previously introduced "resonance" condition
(\ref{resonance}) is indeed the condition for  the appearance of a new
state at the threshold. At the moment of appearance of the new state
the scattering length is:
\begin{equation}\label{ares}
a_{res}=a_0\left(1-i\frac{\cosh(\delta_I)}{\sinh(\delta_I)}\right)
\end{equation}
and its imaginary part attains  its maximum value. In case of weak
absorption ($\delta_I\ll1$)  it can be written as
\begin{equation}\label{aImres}
\mathop{\rm Im}a_{res}=-\frac{a_0}{\delta_I}.
\end{equation}
At the same time, the variation $\Delta a$ due to a small change  of the
real part of the
phase-shift $\Delta \delta_R$  attains its  maximum value:
\[\Delta a=\frac{2a_0\Delta\delta_R}{\cosh(\delta_I)}. \]
Hence  the effect of additional interactions (e.g. the  account of
additional decay channels, strong force, etc.) in the
\emph{near-resonant} situation will result in a fast change of the
\textit{real} part of the scattering length.
  The point at which the $S$-matrix pole crosses the real axis
of the complex $k$-plane can be obtained by comparing formulas
(\ref{zclose}) and (\ref{aImres}):
\[z_{res}=-\frac{\delta_I}{a_0}.\]
It is determined entirely  by the inelasticity parameter. The larger  is
the inelasticity parameter, the larger is the  shift of the $S$-matrix pole
to the left from the origin   of  the complex $k$ -plane,  and the smaller  is
its influence on the scattering cross-sections.

 The energy of the near-threshold state is:
\begin{equation} \label{en}
E_b=\frac{z^2}{2M}=-\frac{\tan^2(\delta_R+\Omega+\pi/8)
}{2Ma_0^2}-\frac{i\delta_I\tan(\delta_R+\Omega+\pi/8)
}{Ma_0^2\cos^2(\delta_R+\Omega+\pi/8)}.
\end{equation}
If the pole $z$ is very close to the real axis of the complex
$k-$plane, so that $\delta_R+\Omega+\pi/8=\pi v+\varphi$ and
$\varphi\ll \delta_I$,  the expression (\ref{en}) should  be
modified:
\begin{equation}\label{ethresh}
E_0=\frac{(-\delta_I+i\varphi)^2}{2Ma_0^2}.
\end{equation}
Here $\varphi$ measures the de-tuning of the phase from the exact
resonance. The state at the threshold is  extremely
extended spatially and has a very small width:
   \[\Gamma=\frac{2\varphi\delta_I}{2Ma_0^2}. \]

   The detailed study of the near-threshold states in $1/R^6$
   potential can be found in \cite{Gao1,Gao2}. In the above cited papers the higher order expansion in $a_0/(a-a_0)$ for the energy of near-threshold state is given:
   \[ E_0=-\frac{1}{2M(a-a_0)^2}\left( 1+\frac{0.918 a_0}{a-a_0}-\frac{0.947 a_0^2}{(a-a_0)^2}\right) \]
   Taking into account (\ref{a6}) we get:
   \begin{equation}\label{Eb4}
   E_0=-\frac{\tan^2(\delta+\Omega+\pi/8)}{2M a_0^2}\left( 1+0.918 \tan(\delta+\Omega+\pi/8)-0.947 \tan^2(\delta+\Omega+\pi/8)\right)
   \end{equation}

We observe that once the scattering length $a$ is fixed
by a certain model,  the
spectrum of near-threshold quasi-molecular states is also fixed (up
to the total number of $H-\bar{H}$ states with given hadron angular
momentum). Indeed, the value of the scattering length uniquely
determines the short-range complex phase-shift $\delta$ and thus the
boundary condition (\ref{delta}), while the solution of the
Schr\"{o}dinger equation with given adiabatic potential and fixed
boundary condition (\ref{delta}) gives
 the spectrum  (up to the number of states). The validity of this statement
is restricted only by
the approximation that our boundary condition (\ref{delta}) is
energy independent, i.e. it is valid for energies $E\ll
\varepsilon_m\approx0.05$ a.u..
We calculate the positions of the  bound states nearest to the threshold,
as they are fixed by  the scattering lengths of
various  models. These  values are collected in Table \ref{Table2}.
We also present the position of the same states when the  spin
dependent strong force is taken into account.
%%%%%%%%%%%%%%%%%%%%%%%%%%%%%%%%%%%%%%
\begin{table}
\centering
\begin{tabular}{|c|c|c|c|c|}
  \hline
                           & Model I    & Model II      & Model III & Model IV \\
  \hline
  $E$, $10^{-5}$ a.u.     &$-10.1-i5.6$& $-2.2-i3.2$ & $-3.8-i13.1$ & $-4.9-i0$ \\
   \hline
  $E^{sc}(S=0)$$10^{-5}$ a.u.&$-8.4-i6.7$ & $-1.4-i3.3$ & $-0.7-i12.8$ & $-4.3-i0.8$ \\
   \hline
  $E^{sc}(S=1)$$10^{-5}$ a.u.&$-7.8-i6.1$ & $-1.3-i2.9$ & $-0.8-i11.5$&$-3.9-i0.6$  \\
  \hline
\end{tabular}
\caption{Energy of the  state nearest to the threshold. The
first row collects the values without account of the strong force,
the second and the third row collects the values with account of the
strong force in spin states $S=0$ and $S=1$ correspondingly.
The  results were obtained
by numerical solution  of the one channel Schr\"odinger equation
subject to the boundary condition given by eq. \ref{delta} with the
phase including  the strong force contribution, together  with  the
square-integrability requirement.}
\label{Table2}
\end{table}
%%%%%%%%%%%%%%%%%%%%%%%%%%%%%%%%%%%%%%%%%%
One can see that the shift of the states due to the account of
nuclear forces is model dependent and is of order $10^{-5}$ a.u. for
the first loosely bound state.  The splitting $\Delta E^{sc}$ of
the triplet and singlet spin states  is on the  order of $10^{-6}$ a.u., which
is one order less than the inelastic width due to rearrangement.
%%%%%%%%%%%%%%%%%%%%%%%%%%%%%%%%%%%%%%%%%%%%%%%%%%%%%%%%%%%%%%%%%
%of the possibility of tuning  the scattering length in the presence
%of external magnetic fields  \cite{Fesh1,Fesh2,Fesh3,Fesh4}.

%With the scattering length $a=6.1-i2.7$ a.u. we find that
%$E_b=-(8.2+i9.1) \cdot 10^{-6}$ a.u. The corresponding lifetime is
%$\tau=7.2$ $10^{-12}$ s.  A  numerical calculation using
%the adiabatic approach in conjunction with the distorted wave approximation
%gives $E_b = -4.6 \cdot 10^{-5}$ a.u. and $\tau = 3.2 \cdot 10^{-12}$ s
%\cite{froe04}.
%From this estimate we find  that
%in the case of $H_{1S}-\bar{H}_{1S}$ the scattering length approximation
%is valid for energies $E \ll E_b\sim 10^{-5}$ a.u.

\begin{figure}
   \centering
  \includegraphics[width=115mm]{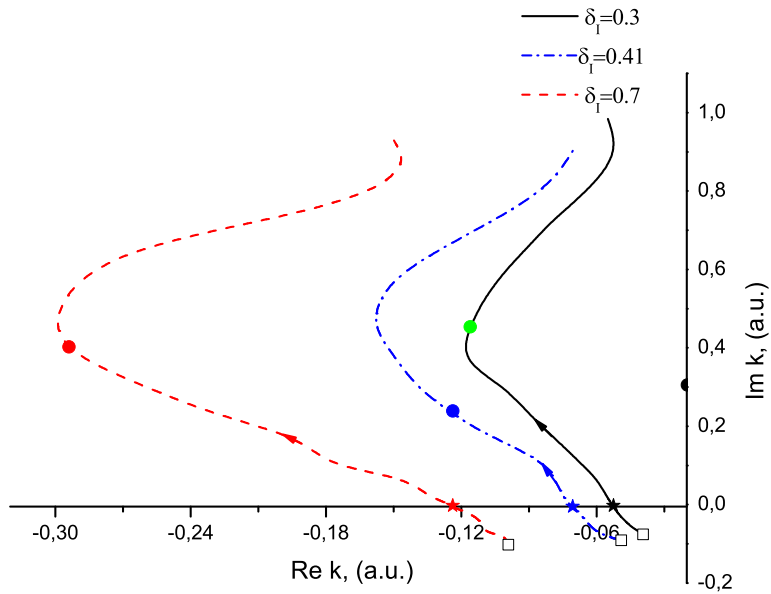}
   \caption{Evolution of the poles of the  $S$-matrix in the complex $k$-plane as function
of the real part of the short-range  phase-shift,  $-\pi/2\leq\delta_R\leq\pi/2$.
Solid  line corresponds to the fixed inelasticity parameter $\delta_I=0.3$ (model I),
 dash-dotted  line - inelasticity parameter $\delta_I=0.41$ (model II)
and dashd line - inelasticity parameter $\delta_I=0.7$ (model III).
 An empty square marks the virtual state, a star marks the point where  the
 $S$-matrix
 pole crosses the real axis of the complex k-plane, a circle - the position
 of the bound state corresponding to the "true" value of the
 phase-shift $\delta_R$, deduced for each model from the corresponding
 scattering length. In calculations, the diverging character of the virtual
states was tempered using the technique of exterior
complex scaling \cite{morg81}.
The arrow shows the direction of the pole evolution
with increasing $\delta_R$.}
    \label{FigPoles}
  \end{figure}

It is useful to see, how the position of the pole that is nearest to the
threshold changes with the short-range phase-shift $\delta$.
Such an evolution of the poles in models I-III is presented on
Fig.\ref{FigPoles}.
%%%%%%%%%%%%%%%%%%%%%%%%%%%%%%%%%%%%%%%%%%%%%%%%%%%%%%%%%%%%%%%%%%%%%%%%
  %%%%%%%%%%%%%%%%%%%%%%%%%%%%%%%%%%%%%%%%%%%%%%%%%%%%%%%%%%%%%%%%%%%%%%%%%%%
 Each of the shown trajectories corresponds to a certain  fixed value of inelasticity parameter
 $\delta_I$.  The  evolution of the
$S$-matrix pole \emph{along the trajectory} can be understood as a
gradual "switching on" the real part of effective interaction,
responsible for rearrangement (while its imaginary part is kept
fixed). At certain value of the phase $\delta_R$ the pole crosses
the real axis of the k-plane, and the new vibrational state
(quasi-bound) appears in the spectrum of $H-\bar{H}$ system. With
increasing the real part of effective interaction (and
correspondingly the phase $\delta_R$) this pole goes far away from
the threshold, becoming deeper bound state. At the same time a new
pole with the negative imaginary momentum (virtual state) is
approaching the real axis. This is reflected in oscillatory behavior
of the elastic and inelastic cross-sections.

We conclude by emphasizing that the resonant dependence of the
cross-sections on the interaction parameters (such as real and
imaginary short-range phase-shift) is of purely quantum mechanical
origin and occurs only in the case of \emph{weak absorption}. This
resonance behavior reflects the existence of the
spectrum of narrow, near-threshold,   quasi-molecular
states of $H-\bar{H}$.

%%%%%%%%%%%%%%%%%%%%%%%%%%%%%%%%%%%%%%%%%%%%%%%%%%%%%%%%%%%%%%%

%%%%%%%%%%%%%%%%%%%%%%%%%%%%%%%%%%%%%%%%%%

\section{Conclusions}

We have presented an overview of the $H-\bar{H}$ collisions based on
the analytical treatment of the asymptotic  potential tail
in conjunction with the
parametrization of the contribution from the inner part of
the interaction.
%Comparing our results with elaborated model
%calculations we establish that in case of hydrogen-antihydrogen
%interaction in \emph{ground state} the value of inelasticity
%parameter, which governs the rate of the transitions to the decay
%channels, turns to be $\delta_I\approx 0.5$.
We show analytically  that for a  realistic value of the
inelasticity parameter $\delta_I$ (the imaginary part of the
short-range scattering phase induced by the rearrangement and the strong-force
interaction)  a  rich spectrum of narrow near-threshold
$H\bar{H}$ states exists.
%The metastability of such states is supported by the long range of
%atom-antiatom interaction and by already mentioned weak absorption
%"in the core".
These states, corresponding to  the near-threshold singularities of
the $S$-matrix, determine the behavior of the elastic and inelastic
cross-sections. The latter  exhibit high sensitivity to even small
perturbation of the $H-\bar{H}$ interaction or its approximate
treatment. We predict important
physical effects such as a  strong isotope effect (i.e. the
resonance-like dependence of elastic and inelastic cross-sections on
the reduced mass of the system) and  demonstrate a significant
dependence of both elastic and inelastic $H-\bar{H}$ cross-sections on
the nuclear potential.

\section{Acknowledgment}

The  research was performed  under support   from the  Wenner-Gren
Foundations, the Royal Swedish Academy of Sciences,
the Swedish  Research Council and the
Russian Foundation for Basic Research grant 02-02-16809.
%A.D. acknowledges the grant  from the
%Chemical Sciences,
%Geosciences and Biosciences of the Office of Basic Energy Sciences,
%Office of Science,
%U.S. Department of Energy.

\vfill \eject

\bibliographystyle{unsrt}
\bibliography{hahresonant}

\vfill \eject

%%%%%%%%%%%%%%%%%%%Effective range %%%%%%%%%%%%%%%%%%%%%%%%%%%%%%%%%%%%

%After eq. (11)

%One can extend the above analysis to the calculation of the
%effective range.
%According to \cite{Gribakin2,Gao1} the
%effective range is:
%\[
%r_e=\frac{a_0}{3}\left[\frac{\Gamma(1/4)}{\Gamma(3/4)}\right]^2\left(1-2\frac{a_0}{a}+2\frac{a_0^2}{a^2}\right).
%\]
%Upon substitution of $a$ from eq. (\ref{a6}) one gets
%\begin{equation}
%\label{effrange}
%r_e=\frac{a_0}{3}\left[\frac{\Gamma(1/4)}{\Gamma(3/4)}\right]^2\frac{(\cot(\pi/8+\Omega+\delta))^2+1}{(\cot(\pi/8+\Omega+\delta)+1)^2}.
%\end{equation}

%Using the effective range, the expansion of $kcot \delta (k)$ can be
%written as
%\begin{equation}
% k cot \delta(k) = - \frac{1}{a} + \frac{1}{2}k^2 r_e
%\end{equation}
%and the improved  cross sections become
%\begin{equation}
%\sigma_{el} = \frac{4\pi} {k^2 + (\frac{1}{2}k^2 r_e - \frac{1}{a})^2}
%\end{equation}
%\begin{equation}
%\sigma_{in} =  ..........................
%\end{equation}

%Now, in the section about the near threshold poles, I suggest a link
%that shows an influence of the  effective range correction on the
%near-threshold pole:

%After eq. (41)

%Including the effective range correction, the binding energy of the
%near-threshold bound states becomes $\frac{k_b^2}{2M}$ with $k_b$
%given by :
%\begin{equation}
%k_b= \frac{1}{a} + \frac{1}{2} k_b^2 r_e
%\end{equation}

\end{document}